\documentclass{JHEP3}

\input{epsf}
\usepackage{epsfig}

\preprint{MIT-CTP-3364 }

 \title{Three-dimensional black hole entropy}
 
\author{ 
Jan Troost and Asato Tsuchiya\footnote{On leave from Department of Physics,
Graduate School of Science, Osaka University, Osaka, Japan.} 
\\      Center for Theoretical Physics \\  MIT \\
    77 Mass Ave \\ Cambridge, MA 02139 \\ USA \\
\email{troost@mit.edu,asato@lns.mit.edu}
   }

 \abstract{ 
We discuss in detail the properties of gravity with a negative cosmological
constant as viewed in Cherns-Simons theory on a line times
a disc. We reanalyze the problem of computing the BTZ entropy,
and show how the demand of unitarity and modular invariance of the boundary
conformal field theory severely constrain
proposals in this framework.}

\begin{document}
\section{Introduction}
The Einstein-Hilbert action in three dimensions, with a negative
cosmological constant (see e.g. \cite{Deser:dr}) can be rewritten in terms of Chern-Simons
theory with gauge group 
$SL(2,R) \times SL(2,R)$ \cite{Achucarro:vz}\cite{Witten:1988hc}.
It is worth studying that Chern-Simons theory in detail
 for several reasons. One reason is that 
gravitational quantum theories on spaces with negative cosmological
constant have been shown to be holographically dual to gauge theories on the boundary,
in the context of string theory. The prime example is the correspondence
between Type IIB string theory on $AdS_5 \times S^5$ which is dual to
$N=4$ super Yang-Mills \cite{Maldacena:1997re}. On the other hand we know that Chern-Simons
theory on $R \times D$ where $D$ is a disk is dual to a chiral
Wess-Zumino-Witten model 
on the boundary\cite{Witten:1988hf}\cite{Elitzur:1989nr}. 
It should be instructive to
study holographic duality in an example that incorporates features
of both theories.

Our focus in this paper will be slightly different though.
We will first of all map out in detail the correpondence between Chern-Simons
states and their spacetime interpretation. We will try to use that knowledge
to recompute the entropy of 
the BTZ black hole \cite{Banados:wn}\cite{Banados:1992gq}.
We attempt in two different
ways to compute the entropy and show that in the non-unitary theory,
there seem to be too many states present to agree with the entropy 
formula. On the other hand, the minimal unitary truncation has too few
degrees of freedom. We thus show how unitarity and the demand of modular
invariance of the BCFT seem to overconstrain the black hole counting
problem in this context.
Our computation differs in crucial respects from those in the literature:
we analyze very concretely the boundary conformal field theory,
 with modular invariant
spectrum, that is induced by the Chern-Simons theory on the boundary
and pay attention to the modular invariance of the proposed spectrum,
as well as to the minimal conformal weight appearing in the theory.

In section \ref{csgravity} we review the Chern-Simons formulation 
of gravity in three dimensions. In the next section, we discuss quantization
of $SL(2,R)$ Chern-Simons theory on a disk with a puncture and map out
the correspondence between classical weights and quantum Hilbert spaces.
Using that correspondence we make a
first connection with BTZ black holes and particle excitations in section \ref{bhandp},
and we discuss a gravitational BPS bound. In sections \ref{btzentropy1}
and \ref{btzentropy2}
we analyze in detail why it is difficult to account for 
 the BTZ entropy in this framework. 

\section{Chern-Simons gravity}
\label{csgravity}
We briefly discuss the Chern-Simons formulation of three-dimensional
gravity with a negative cosmological constant.
We remind the reader that we can write the gravity action in
three dimensions as an $SL(2,R) \times SL(2,R)$
Chern-Simons action (for conventions see appendix \ref{conventions}):
\begin{eqnarray}
S{[}A^+{]}_{CS}-S{[}A^-{]}_{CS} 
&=&  \frac{k}{4 \pi} \int_M Tr \left(A^+dA^+ +\frac{2}{3} {(A^+)}^3\right)
- \frac{k}{4 \pi} \int_M Tr \left(A^-dA^-+\frac{2}{3}{(A^-)}^3\right) \nonumber\\
&=&  \frac{k}{4 \pi l} \int_M \left(e_a \epsilon^{abc} R_{bc}
 + \frac{1}{3 l^2} \epsilon^{abc} e_a e_b e_c\right) +
 \frac{k}{4 \pi l} \int_{\partial M} e^a \omega_a, 
\label{action}
\end{eqnarray}
where $A^{\pm} = \omega \pm \frac{e}{l} $ are two 
$SL(2,R)$ gauge fields and $e$ is the dreibein and $\omega$
is the spin connection one-form. The Newton constant $G$ and the cosmological
constant $\Lambda$ are 
determined by the formulas $k=\frac{l}{4G}$ and $\Lambda=-l^2$, respectively.
Thus the Chern-Simons action is equal to the Einstein-Hilbert action in
three dimensions with cosmological constant
plus a boundary term, which is half the usual Gibbons-Hawking boundary
term (see e.g. \cite{Banados:1998ys}\cite{Rooman:2000zi}).
We will henceforth take the Chern-Simons
theory as the starting point for studying three-dimensional gravity,
without adding any other boundary term.
Hence, the consistent boundary conditions on our fields are
\begin{eqnarray}
\int_{\partial M} Tr (A^{\pm} \delta A^{\pm})=0. 
\label{bc}
\end{eqnarray}

We should warn at the start that there are
fundamental differences between this theory
and our intuition for gravitational
theories.
 The most striking difference between the theory of quantum
gravity that we obtain in this way and our intuition,
 is that in the Chern-Simons theory we allow for any
gauge connection, including the trivial one which would give rise to
a singular geometry in the metric formulation. In other words, the
path integral contains an integral over singular geometries.  
(That is in fact a  feature of the quantum theory that makes 
it renormalizable \cite{Witten:1988hc}.)

Now that we have established the point of view that we will take
on three-dimensional quantum gravity in the Chern-Simons formulation, we make
a substantial digression.
We first study $SL(2,R)$ Chern-Simons theory in some detail to
acquaint the reader with the necessary technical ingredients, without
at first instance
doubling the degrees of freedom because of the product
gauge group $SL(2,R)\times SL(2,R)$ relevant to the gravitational theory.
 After this useful digression,
we will return to recombine the ingredients for the 
product gauge group, i.e. the theory of gravitation.

\section{$SL(2,R)$ Chern-Simons}
\label{sl2r}
\subsection*{Review}
We first fix our conventions for Chern-Simons theory.
We define the Chern-Simons action as:
\begin{eqnarray}
S_{CS}[A] &=& \frac{k}{4 \pi} \int_M Tr \left(AdA+\frac{2}{3}A^3\right).
\end{eqnarray}
We take the trace in the fundamental (Tr) to have normalization
$Tr(T_a T_b) = \frac{1}{2} \eta_{ab}$ where $\eta_{ab}= diag(-1,+1,+1)$
and $a \in {\{} 0,1,2 {\}}$. (For compact
gauge groups $SU(N)$ (where $\eta_{ab}=\delta_{ab}$), this would lead to 
the only consistent
values for $k$ being integers. For $SL(2,R)$ there is no such
restriction at this stage.)

At first, we are
 interested in Chern-Simons theory on a base manifold $M$ that
has the topology of a real line (parametrized by the time $t$) 
times a disk with a
puncture. To the puncture we associate matter in an irreducible
representation of the gauge group.  The Chern-Simons action
 is then supplemented by the particle action $S_p[A,\chi]$:
\begin{eqnarray}
S_{coupled}[A,\chi] &=& S_{CS}[A] + S_p [A,\chi] \nonumber \\
 &=& S_{CS}[A] + \int dt Tr (\lambda \chi^{-1} (t)
(\partial_{t}+A_{t}) \chi(t)). \label{coupled}
\end{eqnarray}
For compact gauge groups, $\lambda$ is a weight for the root system
of the Lie algebra of the gauge group. We will see how this statement
becomes modified for non-compact groups.
We will choose a boundary condition 
\begin{eqnarray}
(lA_t\pm A_{\phi})|_{\partial M}=0,
\label{boundarycondition} 
\end{eqnarray}
which is 
consistent with (\ref{bc}), and we first perform the path-integration over $A_t$,
which gives the following constraint:
\begin{eqnarray}
\frac{k}{2 \pi} F_{12}(t,x^i) + \chi(t) \lambda \chi^{-1}(t) 
\delta^{(2)}(x^i-P) &=& 0,
\end{eqnarray}
where $P$ denotes the puncture in the disk coordinatized by $x^i$
(where $ i \in {\{} 1,2 {\}}$) .
We can solve the constraint in terms of a connection that is
almost pure gauge, and substitute the solution in the action to
obtain the new action 
\begin{eqnarray}
S[U] &=& \frac{k}{4 \pi} \int_{\partial M} Tr \left(U^{-1} \partial_{\phi}
U U^{-1} \left(\partial_{t}\pm \frac{1}{l}\partial_{\phi}\right) U\right) dt d\phi + \frac{k}{12 \pi}
\int_M Tr (U^{-1} dU)^3 \nonumber \\ & & + \frac{1}{2 \pi} \int_{\partial M}
Tr \left(\lambda U^{-1} \left(\partial_{t}\pm\frac{1}{l}\partial_{\phi}\right) U\right).
\end{eqnarray}
The system has reduced to a chiral WZW model on the boundary circle
coupled to a matter source \cite{Elitzur:1989nr}.
For compact groups, we know that when we quantize this system, the
Hilbert space is a representation of the chiral current algebra
which is the Verma module built on the representation of the
gauge group with weight $\lambda-\rho$ (where $\rho$ is half the
sum of the positive roots), modded out by the null vectors.
\subsection*{Extension}
Now we want to analyze how that last statement is modified in $SL(2,R)$
Chern-Simons theory. To study that problem, it is useful to 
remind ourselves of how we can quantize a particle on an $SU(2)$
manifold using the method of orbits. Namely,
we first concentrate on the particle action in (\ref{coupled})
and ignore
the Chern-Simons action and coupling. We can quantize a particle
with resulting quantum spin $j$,
by starting with a classical particle action that is based on the classical
weight $(j +\frac{1}{2}) \alpha$, where $\alpha$ is the single simple
root of $SU(2)$ normalized such that $\alpha^2=2$. 
To show the shift in the weight is non-trivial.
The shift in the weight was first obtained in
 \cite{Nielsen:1987sa}, and we re-obtain it via a
path-integral quantization in appendix \ref{shift} which makes the 
modern treatment
in the paper \cite{Alekseev:vx} more precise.
 In mathematical terms, the shift is fairly
obvious. Indeed, we can build irreducible representations of
$SU(2)$ by quantizing appropriate orbits of $SU(2)$ with the canonical
symplectic form (which we obtain from the decoupled 
particle action). We refer to \cite{Vergne} for a pedagogical discussion.
In that formalism we find that the quantizable (co)adjoint orbits of $SU(2)$
are the two-spheres with  radius $r=\frac{n}{2}$
where $n$ is a strictly positive integer, 
and that the spin of the
representation is indeed related to the sphere of the radius
as $j=\frac{n}{2}-\frac{1}{2}$, thus accounting for all irreducible
representations of $SU(2)$, including the trivial one, using quantizable
orbits of maximal dimension two. 
Luckily this part of the story is readily extendable to $SL(2,R)$.
Indeed, we can obtain all irreducible representations of $SL(2,R)$
that occur in the decomposition of the left regular representation
by quantizing suitable adjoint orbits.
First let us remind the reader of what these irreducible representations
are.

\subsection*{All unitary irreducible representations of  $SL(2,R)$}
\begin{figure} 
 \epsfxsize=14cm
\epsfbox{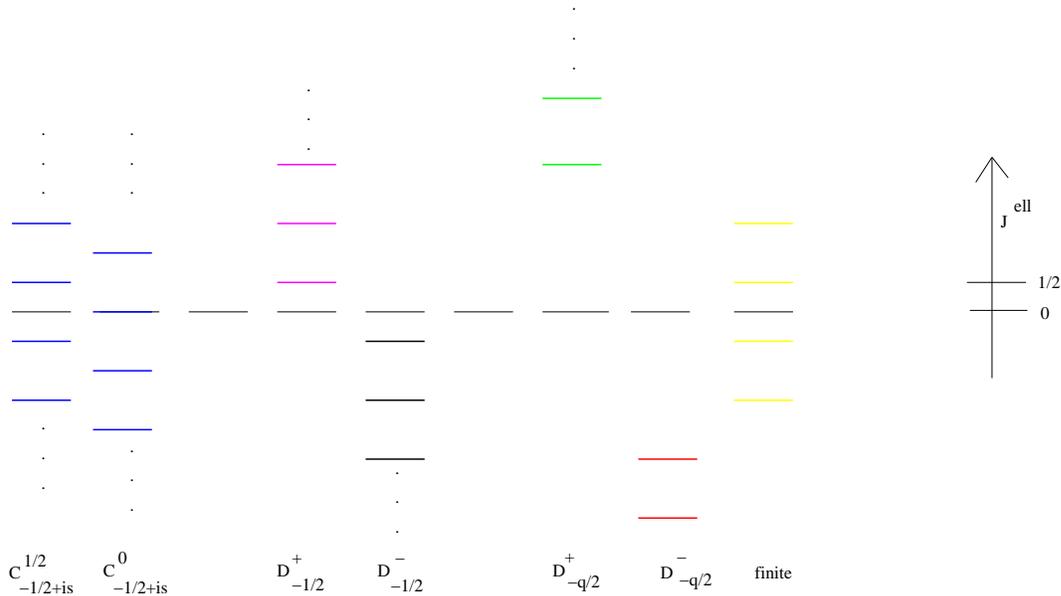}
\caption{\em Spectrum of an elliptic operator \em \label{sl2rladders}}
\end{figure}
We give a precise classification of all unitary irreducible
representations of $SL(2,R)$ following \cite{Lang} \cite{Klimyk}.
The unitary irreps are:
\begin{itemize}
\item The continuous representations. There are two series of continuous
representations and the series are
labeled by the discrete parameter $\epsilon \in {\{} 0,\frac{1}{2} {\}}$
which indicates the parity of the representation.
We denote them by $C^{\epsilon}_{-\frac{1}{2}+i s}$ where
$s$ is real and the quadratic Casimir $c_2$ takes the
values $c_2 = \frac{1}{4}+s^2$ (or $c_2 = -\tau(\tau+1)$ where
$\tau=-\frac{1}{2}+i s$ in perhaps more familiar notation).
These representations have a spectrum for a normalized elliptic generator
$J^{ell}= T^0$ that is either all integers, for $\epsilon=0$, or all
half-integers, for $\epsilon=\frac{1}{2}$. (See figure \ref{sl2rladders}.)
\item The true discrete representations. There are again two 
series, which are lowest weight representations and highest
weight representations. These are denoted $D^{+}_{\tau}$ or
$D^{-}_{\tau}$ respectively. The index $\tau$ takes the
values $\tau \in {\{} -1, -\frac{3}{2}, -2 ,\dots {\}}$, and 
the quadratic Casimir is again $c_2= -\tau (\tau+1)$.
The spectrum for the elliptic generator is given by $-\tau+n$
where $n$ is a positive integer (or zero) for the $D^{+}_{\tau}$
representation and $\tau-n$ for the $D^{-}_{\tau}$ representation.
(See figure \ref{sl2rladders}.)
\item The mock discrete representations. There are two of these,
namely $D^{\pm}_{-\frac{1}{2}}$ with spectra similar to the
ones described for the true discrete series.
(See figure \ref{sl2rladders}.)
\item The  complementary representations. These have $-1<\tau<0$
and $\tau \neq \frac{1}{2}$. They are of even parity. 
\end{itemize}
We make two remarks here. First of all, to the true discrete series
with index $\tau$ 
there is naturally associated a finite dimensional non-unitary
representation of dimension $-2 \tau-1$ with elliptic eigenvalues
that lie between the eigenvalues of the two discrete representations 
\cite{Lang}.
(See last column in figure \ref{sl2rladders}.)
The second remark is that the mock discrete series are the only discrete
representations that naturally combine to form a continuous representation
as is apparent from the spectrum for the elliptic generator.

Now, the left regular representation on quadratically integrable 
functions on the group manifold $SL(2,R)$ decomposes into 
true discrete representations, and both series of 
continuous representations
(with a known Plancherel measure \cite{Klimyk}). These representations only
we should be able to obtain by quantizing suitable orbits
\cite{Vergne}.

The detailed proof of that statement is lengthy. We refer the reader
to \cite{Vergne} and just summarize our intuition. First of all,
let's  draw a picture of the orbits of $SL(2,R)$
(see figure {\ref{conjugacyclasses}). The structure of the orbits
is easily understood when
we realize that $SL(2,R)$ is isomorphic to the Lorentz group in three dimensions
$SO(2,1)$.
\begin{figure} 
 \epsfxsize=10cm
\epsfbox{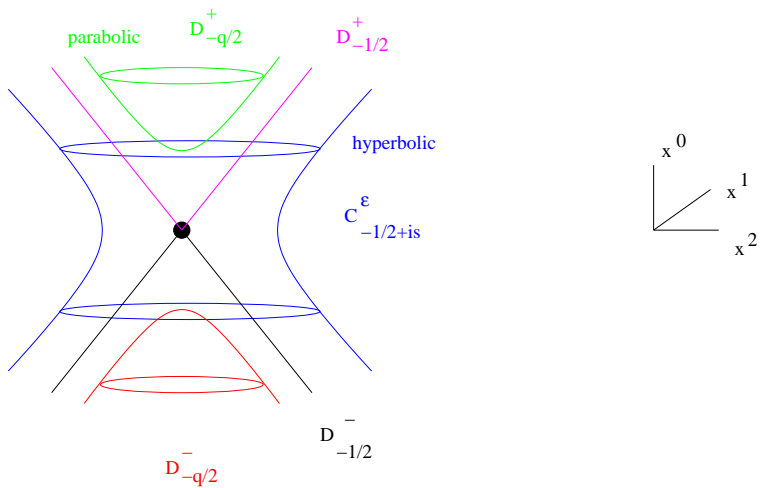}
\caption{\em Conjugacy classes of $SL(2,R)$ \em \label{conjugacyclasses}}
\end{figure}
We notice that the conjugacy classes of $SL(2,R)$ are of several
kinds: a paraboloid above the $(x^1,x^2)$ plane, one
below the $(x^1,x^2)$ plane, a hyperboloid associated 
with nearest approach  to the $x^0$-axis, the point at the origin,
and a future and past light-cone with the point at the origin
removed. Note that the hyperboloid, as we move it towards the origin,
when it reaches the origin splits into two light-cone sheets and evolves
further into paraboloids.

The correspondence between orbits and representations goes as
follows (see also \cite{Witten:1987ty}). The hyperboloids of radius $s$ correspond to 
two different continuous representations  
$C_{-\frac{1}{2} + is}^{0,\frac{1}{2}}$ -- we have two ways
of quantizing each orbit. When we pass the hyperboloid through
the origin, it splits into two paraboloids. Thereafter, the
quantizable orbits are the paraboloids with half-integer radius $r$, which correspond to
the discrete representations  
$D^{+}_{\tau=-r-\frac{1}{2}}$
and
$D^{-}_{\tau=-r-\frac{1}{2}}$ depending on whether we take
the upper or the lower sheet.\footnote{We can think of the doublesheeted
hyperboloid as splitting not only into two paraboloids, but also
as obtaining radii which differ by a half-integer, since the
spectrum of the discrete representations is also either half-integer
or integer depending on the value of the radius.}
That completes the map of regular representations and quantizable
orbits. (Note that the method of orbits gives good intuition
for the spectrum of the elliptic operator in a representation associated
to a given geometrical orbit. To convince oneself of this fact, the
reader should merely try to match up figure \ref{conjugacyclasses}
with figure \ref{sl2rladders}, which is a useful mental exercise.)

But, from the geometrical picture of orbits, another representation
theoretic fact becomes intuitive. The orbits corresponding to
continuous representations split up into two light-cones plus 
the point at the origin when we let $s$ approach $0$. We thus
gain intuition for the fact that the continuous
representation with $s=0$ (and $\epsilon=1/2$) decomposes as $D^{\pm}_{-\frac{1}{2}}$,
i.e. into two mock discrete representations.\footnote{
Moreover
the gap that arises when we continue further, between the tips of
the parabolas corresponding to 
the discrete representations $D^{\pm}_{\tau}$ is a measure for
the size of the non-unitary finite dimensional representations that
is naturally associated to these discrete representations. We can 
picture a ball between the tip of the parabolas corresponding to
the well-known spherical orbits of $SU(2)$ and their corresponding
finite dimensional representations, after analytic continuation --
if we really insist on a pictorial representation.}
Strictly speaking though, the mock discrete representations
are not obtained by the orbit method, nor are the complementary
representations.

\subsection*{Conclusion}
We take away from these intuitive mathematical facts the following
statements about the quantization of a particle with an action
determined by a weight in these orbits. When the particle has
hyperbolic weight, the associated representation is continuous.
When it has non-zero weight on the light-cone, the only invariant under conjugation
is the overall sign of the weight. The sign determines
whether the associated unitary Hilbert space is $D^{+}_{-\frac{1}{2}}$
or $D^{-}_{-\frac{1}{2}}$. When the weight is zero, the representation
is trivial. When the weight is elliptic, it is quantized, and the
particle Hilbert space will be a discrete representation. 
(See table \ref{correspondencetable}.)
It would be useful to back up these statements with a path integral
computation for a particle on an $SL(2,R)$ manifold, but we refrain
from carrying out this exercise in this paper.
\begin{table}
\begin{tabular}{||c||c|c|c|c|c||}
  \hline         
   $\lambda $  &  $\pm (-q+1) T_0 $ and $q \ge 2$ integer  & $2 s T_2$ & $\pm(T_0+T_2)$  & 0 & - \\ 
\hline
irrep   & $D^{\pm}_{-\frac{q}{2}}$ & 
$C^{0,\frac{1}{2}}_{-\frac{1}{2}+is}$ &$ D_{-\frac{1}{2}}^{\pm}$ &
trivial & complementary  \\
\hline
\end{tabular}
\caption{\em The correspondence between classical (generalized) weights
in the particle action and irreducible quantum Hilbert spaces.
\label{correspondencetable}}
\end{table}

The complementary representations are not obtained
by the orbit method. They lie in the far quantum regime of the
particle action, since the radius of the orbit effectively acts
like the particle action coupling constant \cite{Witten:1987ty}. The complementary
representations clearly are associated to small Casimirs. So,
although we have no classical action starting point to study
particles in complementary representations, since they are unitary
we can accept them as quantum mechanical representation spaces.
\footnote{In an intuitive sense, these representations exactly appear to fill the
hole left by the quantum shift in the classical weight. }

In summary, by reviewing the orbit method for obtaining irreducible
representations of $SL(2,R)$, we have hopefully convinced the reader
of the fact that we are able to quantize the $SL(2,R)$ particle action.
We have laid out the detailed map between classical weights in the
action (\ref{coupled}) and quantum particle Hilbert spaces (see table
\ref{correspondencetable}).

\subsection*{Unitarity and current algebra}
When we return to quantizing the full action, including the 
Chern-Simons term that gives rise to a chiral WZW model on the
boundary, we run into subtleties. A naive quantization of the
model will give rise to the usual representation of the current
algebra on the irreducible representation of $SL(2,R)$.
The model will be non-unitary, because of negative norm states 
arising from lowering operators associated to the elliptic
one-parameter subgroup of $SL(2,R)$ (i.e. ``the raising operators
associated to the time-like direction''). 

So, we can quantize a particle on $SL(2,R)$ by analogy to a particle
on $SU(2)$ consistently. But, when we continue the analogy naively
to include the action of the current algebra, we arrive at a 
non-unitary theory. We note at this stage that at
 least for a bulk without a boundary, there exists
a quantization of Chern-Simons theory with non-compact gauge group
which is unitary
\cite{Bar-Natan:1991rn}. It would be interesting to figure out whether
there is a close
analogue of the quantization for the theory on the disk, 
and if possible, to understand the
relevant unitary representation of the current algebra. We will propose
to solve
the problem of unitarity in two different ways later on.

In passing we note that when we quantize strings on $AdS_3
\times S^3 \times T^4$, for example, then the non-unitarity of the
representations of the $SL(2,R)$ loop group is not a problem after determining
the right spectrum, since the ghosts
decouple after applying the appropriate constraints (or after computing
the BRST cohomology) \cite{Maldacena:2000hw}. 

Finally we make a remark on winding sectors. We have not been careful in 
treating the non-trivial topology of the gauge group with 
first homotopy group $\Pi_1(SL(2,R))=Z$, since it was not crucial for
our purposes until now. Later on, we will need the fact that the representation
of the current algebra built on the discrete representations satisfies the
following relation: $\hat{D}^{+,w}_{-\frac{q}{2}} \equiv \hat{D}^{-,w-1}_{\frac{q}{2}-\frac{k}{2}}$
which was proved in \cite{Maldacena:2000hw}. We could derive it in the Chern-Simons
theory by studying the transformation of the Hilbert space when we move from a 
sector with winding number $w$ to a sector with winding number $w-1$.

\section{Black holes and particle excitations}
\label{bhandp}
We have reminded ourselves of useful properties of $SL(2,R)$ Chern-Simons
theory, and we will now bring them to bare on understanding three-dimensional
gravity with a negative cosmological constant, with signature $(-1,+1,+1)$,
which can be rewritten as a $SL(2,R) \times SL(2,R)$ Chern-Simons
theory as reviewed in section 2. In particular, we will be able to associate holonomies to
classical gravity solutions, and we can thus associate them to 
punctures with particular $SL(2,R)$ representations in the Chern-Simons
theory. We also  clarify the role of the massless BTZ black
hole solution, of the $AdS_3$ solution, and of the trivial Chern-Simons
background.
 
Let's study
some classical solutions to the equations of motion with a source term.
We restrict to the BTZ black hole solutions and compute the 
corresponding dreibein and spin connection, next to compute the
holonomy of the classical gauge field around the source. That
provides us with a map between classical sources and holonomies.
Next, we can look at the quantum theory with the knowledge acquired
in section \ref{sl2r}.

To make contact with the usual metric formulation of three-dimensional
gravity, we start out with the familiar BTZ 
metric\cite{Banados:wn}\cite{Banados:1992gq}:
\begin{eqnarray}
ds^2 &=& -\left(-M+\frac{r^2}{l^2}+ \frac{J^2}{4 r^2}\right) dt^2
+ \left(-M+\frac{r^2}{l^2}+ \frac{J^2}{4 r^2}\right)^{-1} dr^2
\nonumber \\ & &
+ r^2 \left(d \phi - \frac{J}{2 r^2} dt\right)^2
\label{genmet}
\end{eqnarray}
where $\phi \in {[}0, 2 \pi {[}$.
The mass $M$ and angular momentum $J$ of the black hole are expressed
in terms of the outer and inner horizon $r_{\pm}$ as 
$M = \frac{r^2_+ + r^2_-}{l^2}$ and $J= \pm \frac{2 r_+ r_-}{l}$.
The inner and outer horizon are then located at the (positive)
square root of:
\begin{eqnarray}
r^2_{\pm} &=& \frac{M l^2}{2} \left(1 \pm \sqrt{1-\left(\frac{J}{Ml}\right)^2}\right).
\end{eqnarray}
We choose the following dreibein:
\begin{eqnarray}
e^0 &=& -f dt \nonumber \\
e^1 &=& f^{-1} dr \nonumber \\
e^2 &=& r d \phi - \frac{J}{2r} dt
\end{eqnarray}
where $f^2 = r^2/l^2 -M+J^2/(4 r^2)$.
Labeling $(t,r,\phi)$ tangent directions as $(0,1,2)$ respectively,
we obtain the one-form:
\begin{eqnarray}
\omega_{01} &=& \frac{r}{l^2}dt-\frac{J}{2r}d\phi \nonumber \\
\omega_{02} &=& - \frac{J}{2 r^2} f^{-1}dr \nonumber \\
\omega_{12} &=& -f d\phi.
\end{eqnarray}
{From} these we easily get the gauge potentials $A^{\pm}$, and see that those 
potentials satisfy our boundary conditions (\ref{boundarycondition}) at $r=\infty$:
$l A^{\pm}_{t}\pm A^{\pm}_{\phi}=0$  holds for arbitrary $r$.
We can compute
the gauge invariant expectation value of a Wilson loop looping the origin,
and we find \cite{Cangemi:1992my}\cite{Carlip:1995qv}:
\begin{eqnarray}
Tr_F e^{\oint A^{\pm}} &=& 2 \cosh \left(\pi \sqrt{M\pm\frac{J}{l}}\right).
\end{eqnarray}
Now we interpret the formula. First of all, it is clear that the non-trivial
black hole solutions correspond to non-trivial source terms for the 
equations of motion. The holonomy is non-trivial  when the field
strength is sourced inside the (otherwise) topologically trivial Wilson
loop. Thus, we can think of a spatial section of the BTZ space-time 
as a disk with a puncture where a source is inserted. That establishes
the topology of a BTZ space-time.

Next, we notice that a generic BTZ black hole is associated to 
holonomies that are greater than two, i.e. they correspond to 
hyperbolic orbits. These orbits
are naturally associated to continuous representations
of both the left and right $SL(2,R)$ symmetry group.
By comparing the holonomy that we calculated this way to the
source term in the coupled Chern-Simons and particle action,
we can associate the weight $\frac{2 \pi}{k} \lambda^{\pm}  
= 2 \pi  \sqrt{M \pm J/l} T_2$ or $ \lambda^{\pm} = k \sqrt{M \pm J/l} T_2$
to a black hole with mass $M$ and angular momentum $J$.
It is thus associated to a continuous representation with values
$s^{\pm}=\frac{k}{2}  \sqrt{M \pm J/l}$.

Next, we analytically continue the formula for the holonomies
to negative $M \pm J$ and find:
\begin{eqnarray}
Tr_F e^{\oint A^{\pm}} &=& 2 \cos \left(\pi \sqrt{-M \mp \frac{J}{l}}\right).
\end{eqnarray}
Following the same reasoning as before, we
 find the relevant weight for the quantum Hilbert space:
 $\lambda^{\pm} =  \mp k \sqrt{-M \mp  J/l} T_0$.
We know that when we study the true $SL(2,R)$ group (and not its
covering), the value of $\lambda$ is quantized as
$\lambda^{\pm} =  \pm (-q+1) T_0 =  \mp k \sqrt{-M \pm J/l} T_0 \;\;(q\geq 2)$. We thus find
that we can only quantize space-times with $ k \sqrt{-M \pm J/l} = q-1$
in that Chern-Simons theory. We notice for instance that the $AdS_3$
geometry (-- not the covering --) with $J=0$ and $M=-1$ is
only quantizable when $k$ is an integer. This is one way to realize
that integer $k$ do play a special role in the $SL(2,R)$ theory.\footnote{
We note in passing that other even more indirect
ways to realize  the special role of integer
levels is by noticing the spectral flow
relation,  by identifying the level of the $SL(2,R)$
theory with quantized charges in string theory, or by trying to combine
an $SL(2,R)$ WZW model with a compact WZW model to obtain a supersymmetric
theory.}

Now we address a more subtle point. We notice that a light-cone
value for the holonomy, after taking the trace, always gives
rise to the trivial value (i.e. two) for the Wilson line. Nevertheless,
we know that we can have non-trivial source terms in the classical
equation of motion which correspond to light-cone weights. And we
know that this statement is gauge invariant (since the light-cones
are invariant under Lorentz transformations). 
Thus, for $M=0=J$, we have a special situation, with trivial value
for the Wilson loop, but with three different classes of gauge fields which
are gauge inequivalent. One class is the trivial gauge connection,
associated to the trivial representation. This corresponds to the
unbroken vacuum of the Chern-Simons theory. The massless BTZ black hole
is non-trivial, and can be associated to the $D^{+}_{-\frac{1}{2}}$
representation, while the massless BTZ black hole with negative
values for the radial variable $r$ can be associated to 
$D^{-}_{-\frac{1}{2}}$. We note that this is not unexpected in view
of the fact that on the Poincare patch (AdS) time flows in opposite
directions when we continue through $r=0$, and this is reflected
in the fact that the spectrum for the elliptic operator flips sign.
(Of course, a lot of these properties are foreshadowed in the
non-trivial geometry of the three-dimensional black hole \cite{Banados:1992gq}.)

An extremal black hole corresponds to a Wilson loop that is trivial
(in the sense that it is equal to the Wilson loop of a gauge field
that is zero) in either the left or the right sector. It is then
associated to a product of a continuous representation and
a trivial, or a $D^{\pm}_{-\frac{1}{2}}$ representation,
depending on the precise choice of the sign of the associated
light-cone weight.

We might also want to study complementary representations
of $SL(2,R)$, i.e. the only remaining unitary representations. 
These representations can  be suitably combined, leading
presumably to particle like excitations in $AdS_3$ (with non-trivial mass and
spin). It
would be interesting to further study the possible geometric interpretations
of more general combinations of $SL(2,R) \times SL(2,R)$ representations.
We only note at this point that if we combine
complementary or discrete with continuous representations, 
we violate the bound $|M| \ge |J|$.

\subsection*{Gravitational BPS bound}
Let us digress at this point and discuss the gravitational
BPS-like bound. 
We can think of the bound  $|M| \ge |J|$ as a BPS bound
arising from a supersymmetric version of the Chern-Simons theory
that we are studying
(see e.g. \cite{Izquierdo:1994jz}).
Note that we  actually should be able to study the quantized
theory in this context, and to derive the bound from the action of
quantum supercharges on a Hilbert space, in analogy to the derivation
in supersymmetric field theory. Imagine we can derive this bound
in the quantum theory (for instance by studying the representation
theory of the boundary superconformal algebra). 
Then, it turns out that the quantum gravitational
BPS bound
is classically interpreted for the BTZ black holes (with $M \ge 0$)
as the condition for absence of a naked singularity, while
for $M<0$, it is interpreted as implying the absence of closed
time-like curves. Thus, in the classical bosonic theory the BPS
bound arises from physical considerations, while in the supersymmetric
quantum theory, we imagine it arising through representation theory.
Working out precisely these conceptual connections in the quantum 
theory will be worthwhile.

\section{BTZ entropy I}
\label{btzentropy1}
In this section, we  refine the picture described above,
and we  try to compute the BTZ entropy in the non-unitary theory
of quantum gravity described in the previous sections. After analyzing
in detail why the computation fails, we formulate a new proposal for
a unitary boundary conformal field theory 
in the next section.
However, we will find that unitarity 
seems to overconstrain
the black hole counting problem.

We attempt to compute the BTZ entropy as follows.  First, we need to
identify which states in Chern-Simons theory on a disk with a source
correspond to a BTZ black hole. The generator of time translations
$L_0+\bar{L}_0$ in the boundary theory is the Hamiltonian,
and determines the mass of the full system (including boundary
excitations), which we will identify
with the mass of the space-time. Similarly, $L_0-\bar{L}_0$
corresponds to the total angular momentum. Properly normalizing the Sugawara
energy momentum yields (in the semi-classical limit):
\begin{eqnarray}
L_0 &=& \frac{1}{k-2} :J^a_{-n} J_{a,n}: = \frac{1}{16G} ( lM+J) \nonumber \\
\bar{L}_0 &=&  \frac{1}{k-2} :\bar{J}^a_{-n} \bar{J}_{a,n}: = \frac{1}{16G} ( lM - J).
\end{eqnarray}
It is now easy to correct our statements in the previous sections.
In section \ref{bhandp} we associated a particular mass and angular
momentum to a primary state of the boundary conformal field theory.
I.e. the boundary conformal field theory had no oscillator excitations
turned on. When we turn on these excitations, we obtain states
with a higher total mass. The total mass of the system should be
identified as the mass of the space-time. (And analogously
for the angular momentum.) Thus, there are many states, built on
different ground states, which have the same total value for
the mass. The idea behind the black hole entropy counting argument
is that we count the number of states corresponding to a given
total mass and angular momentum (and fixed topology). We will try to count these
states in the semi-classical limit, where the Bekenstein-Hawking
entropy formula is valid, i.e. in the limit $l >> G$ or $k >>1$.
The formulas in this section should be understood to be  valid
in this limit only.

To be able to compute that number of states easily, we need
a couple of non-trivial ingredients: the bare central charge,
the minimal conformal dimension in the theory, and a modular
invariant spectrum. The bare central charge of the theory is
easy to determine: it is $c=\frac{3k}{k-2}$. (For the connection
to the standard result of \cite{Brown:nw} see \cite{Banados:tn}.) The minimal conformal
dimension in the theory depends on the spectrum of the theory,
which we want to choose in a way consistent with modular invariance.
A reasonable proposal for the spectrum seems to be to take the
following current algebra representations in the Hilbert space \cite{Maldacena:2000hw}:
\begin{eqnarray}
\hat{D}^{+,w}_{-\frac{q}{2}} & \otimes & \hat{D}^{+,w}_{-\frac{q}{2}} \nonumber \\
\hat{C}^{\alpha,w}_{-1/2+is} & \otimes & \hat{C}^{\alpha,w}_{-1/2+is}
\end{eqnarray}
where $-\frac{k-1}{2}<-\frac{q}{2}<-\frac{1}{2}$ and $w$ is a winding number that labels
different sectors of the theory. We also need to specify
a measure for the continuous representations which will not be
crucial for our purposes. We will discuss subtleties associated
to this choice of spectrum a little later.

Given the spectrum, we can determine the minimal conformal dimension
and compute the effective central charge of the theory. In the zero
winding sector, the minimal conformal dimension is (in the semi-classical
limit) $\Delta_{min}^{w=0}= -\frac{k}{4}$, which would give an attractive
central charge of $c_{eff}=c-24 \Delta_{min} = 6k = \frac{3l}{2G}$. In turn, that would lead, 
following the elegant argument of \cite{Strominger:1997eq}
 to a swift derivation of the desired BTZ black hole entropy.
Unfortunately, this story is not convincing for the following reasons.

\subsection*{Critique}
When we analyze the minimal conformal weight in the winding sectors,
we find that the conformal weights are not bounded from below. That
undermines fatally our first attempt. Note that there would be other valid
critiques of the derivation.

One issue that we need to address is the precise modular invariant partition function of the
$SL(2,R)$ Wess-Zumino-Witten theory. In fact, in the literature we find strong arguments
for the proposed spectrum \cite{Maldacena:2000hw} (see also the important footnotes
5, 6 and 18 in \cite{Maldacena:2000hw}), we find the computation of the
free energy of string theory on $AdS_3 \times N$, which yields a modular invariant
\cite{Maldacena:2000kv} result, and there is moreover an analysis of the factorization
of four-point functions in the relevant analytically continued conformal field 
theory \cite{Maldacena:2001km}. 
But it seems that the modular invariant partition function
has not been written down as a simple and clearcut formula (see also the footnote 2 in
\cite{Maldacena:2000kv}). What does seem clear is that the winding sectors are crucial
to obtain a modular invariant spectrum. That seems sufficiently devastating.
 
The second most striking shortcoming of our analysis (apart from the bottomless spectrum)
is that the boundary theory is non-unitary because states with negative norm
can be created by applying appropriate creation operators to primary states. (The 
representations for the zero-modes are unitary.) To mend this shortcoming, one
 could try to reproduce the argument
in the unitary $SL(2,C)/SU(2)$ conformal field theory, and the associated Chern-Simons
theory (see e.g. \cite{Witten:1989ip}), 
in  other words, in a euclidean setting. The fact that the four-point functions
of the euclidean conformal field theory \cite{Teschner:1997ft} factorize over, amongst others, short string states 
(i.e. states in discrete representations) then could become an important part of completing
the  argument.
 
\section{BTZ entropy II}
\label{btzentropy2}
We obtained a non-unitary boundary theory, where modular invariance of the boundary partition
function forced a fatal result for the minimal conformal weight. In this section we propose
a modified theory which is unitary, and has a clearcut modular invariant partition function
with minimal conformal weight. We will  show in detail why this modified theory also does
not give rise to the right counting of degrees of freedom.

The argument runs as follows.
To obtain a unitary theory, we have to get rid off the modes that give rise to negative norm
states (at least). The minimal way to achieve this, 
for Chern-Simons theory on a disk, is to add
a boundary coupling. We choose the boundary 
coupling to be an interaction term that
includes a boundary gauge field that gauges 
the elliptic $U(1)$ subgroup of $SL(2,R)$ at
the boundary. 
We thus leave the bulk gravitational theory intact, and modify the action by
a boundary term only. 
The resulting boundary model will be the coset two-dimensional conformal
field theory $SL(2,R)/U(1)$. The resulting gravitational theory 
has a space-time Hamiltonian which is
the coset conformal field theory Hamiltonian 
(since that is the operator that generates time
translations), and a similar reasoning holds once again for
the angular momentum. We obtain:
\begin{eqnarray}
L_0^{cs} &=& L_0^{SL(2,R)}- L_0^{U(1)} = \frac{k}{4}\left(M+\frac{J}{l}\right) \nonumber \\
 \bar{L}_0^{cs} &=& \bar{L}_0^{SL(2,R)}-\bar{L}_0^{U(1)} = \frac{k}{4}\left(M-\frac{J}{l}\right) \nonumber \
\end{eqnarray}
where the allowed states are the ones with $J^{\pm}_{-n}$ excitations and with 
$J_0^0-\bar{J}^0_0=n$ and $J_0^0+\bar{J}^0_0=-kw$.
The numbers $n$ and $w$ are $U(1)$ 
charges under the unbroken $U(1)$ subgroups of $SL(2,R) \times SL(2,R)$,
and in the context of the cigar CFT they are associated to non-trivial
winding and momentum in the angular direction \cite{Dijkgraaf:1991ba}.

Let us discuss a few aspects of our proposal for a modified, unitary theory.
The truncation of the boundary theory might seem arbitrary. We argue that it
is the minimal truncation that preserves unitarity. 
The original theory in fact contains ghosts
and can intuitively be argued to have too many states 
to obtain the correct entropy law. The truncation we proposed 
above is the one that precisely only gets rid of the states responsible
for non-unitarity. 
At the same time, we note that the requirement of a precise modular invariant 
partition function
and spectrum also constitute an argument for the proposed truncation. 
If we desire the boundary 
conformal field theory to be consistent, a spectrum giving rise to 
modular invariance seems imperative.

The truncation moreover seems to interfere with perhaps desirable
gluing properties of Chern-Simons theory. 
It may be thus useful to remark that our truncation may be
achieved in another manner. We could include in our action a bulk Chern-Simons $U(1)\times U(1)$ action,
and include a ghost boundary term, and moreover demand that our boundary states are in a given
BRST cohomology (see \cite{Dijkgraaf:1991ba}).
 That will lead, in the particular example of the theory on the disk to the same boundary
action and boundary theory. The decoupled bulk equations still allow for the standard gravity solutions.

But, in any case, the truncation gives rise to a special role for the $U(1) \times U(1)$ symmetry that we gauged. In fact,
we are left with only a $U(1) \times U(1)$ symmetry (which conformal field theorists may think off as associated to the 
angular direction of the semi-infinite cigar or two-dimensional black hole) which in our context
is the charge of two components of the dreibein or of the spin connection under a rigid rotation.
The assignment of these charges and their incorporation in the spectrum are crucial for modular
invariance. The fact that our fundamental variables are the dreibein and the spin
connection is thus important. (In this respect, our formalism is reminiscent of loop quantum gravity.)

We note also that the demand of modular invariance incorporates other 
non-trivial information
(compare e.g. to \cite{Henneaux:1999ib}) as seen below.
It implies that the two chiral sectors of the theory are linked.
Although the two sectors are
already related by the demand of quantization of space-time angular momentum, 
our insistence
on a modular invariant spectrum also locks the zero-modes of the two chiral 
sectors, and chooses
to combine the left and right spectrum in a very particular way.

Now let us compute the black hole entropy in this unitary theory.
In this theory the argument for black hole state counting 
can be made precise, but it does
not lead to the expected result. The bare
central charge is $c=\frac{3k}{k-2}-1$, and the spectrum of states 
needed for an explicitly
modular invariant partition function is given by \cite{Hanany:2002ev}:
\begin{eqnarray}
\hat{D}^{+}_{-\frac{q}{2}} & \otimes & \hat{D}^{+}_{-\frac{q}{2}} \nonumber \\
\hat{C}^{\epsilon}_{-1/2+is} & \otimes & \hat{C}^{\epsilon}_{-1/2+is}
\end{eqnarray}
where $-\frac{k-1}{2} <-\frac{q}{2}< -\frac{1}{2}$  and $J^0_0$ and $\bar{J}^0_0$ satisfy the constraints
$J_0^0-\bar{J}^0_0=n$ and $J_0^0+\bar{J}^0_0=-kw$.
The conformal weights of the primaries are given by:
\begin{eqnarray}
h_{primary} &=& -\frac{q(q-2)}{4(k-2)}+ \frac{(n-kw)^2}{4k} \nonumber \\
\bar{h}_{primary} &=& -\frac{q(q-2)}{4(k-2)}+ \frac{(n+kw)^2}{4k}. 
\label{primaryweight}
\end{eqnarray}
Here we see explicitly
that the space-time angular momentum is quantized.
To determine the minimal conformal weight, we need to take into account the spectrum 
of the elliptic generator in the discrete lowest weight representations (as
discussed in section \ref{sl2r}), and we find 
that the $U(1)$ contribution to the
conformal weight (i.e. the second term in formula (\ref{primaryweight})) 
forces a minimal conformal weight that is positive, consistent
with the unitarity of the cigar conformal field theory.\footnote{We thank Juan
Maldacena for pointing out a crucial oversight on this point in the original version of our paper
which lead to a faulty conclusion.}
The effective
central charge is then too small to be able to account for the expected black hole
entropy.

In summary, 
we  analyzed in detail two proposals for boundary conformal field theories, and showed that both in the
non-unitary theory and in the unitary theory, the demand of modular invariance for the boundary
partition function forced a fatal result for the effective central charge. It seems that even a minimal unitary
truncation of the unitary theory does not allow for a sufficient number of degrees of freedom to account
for the black hole entropy. Indeed, in every unitary framework, the effective central charge will not be larger
then the bare central charge. Although one would like to appeal to a negative conformal weight for an
$AdS_3$ ground state to raise the effective central charge, that is difficult to
reconcile with unitarity. 

Some comments on our analysis are in order.
Our analysis should be compared to those in the literature.
(see e.g.
\cite{Carlip:1994gy,Carlip:1996yb,Strominger:1997eq,Birmingham:1998jt,Banados:1998ta,Fernando:2000kv,
Govindarajan:2001ee,Fjelstad:2001je}).
For a critique of the approaches in \cite{Carlip:1994gy}\cite{Carlip:1996yb}\cite{Strominger:1997eq}
we refer to the paper \cite{Carlip:1998qw}.
The main difference between these papers and ours is that we explicitly
identify proposals for the dual conformal field theory, and the degrees of freedom that should be
responsible for the black hole
entropy.

There has been some discussion in the literature of what the relevant 
degrees of freedom
are, and where they are located. In our  theory they are clearly the degrees
of freedom located at the boundary of the manifold 
(and associated to the punctures on the disk). 
The boundary  can be viewed as asymptotic infinity,
as it would in an  approach as in \cite{Strominger:1997eq}, 
or as located on the horizon, as in \cite{Carlip:1999cy}. From our 
perspective, we
need to identify the full system, both the punctured disk and the boundary 
degrees of freedom 
as describing a specific black hole state. 
The total mass of the system, which we identify with the
black hole mass, is given by both contributions from the puncture 
(the weight of the primary) and
the excitations at the boundary. From this viewpoint, 
it may perhaps be more natural to think of
our system as describing the interior of a black hole, 
say the space beyond the event horizon.
If these statements seem counterintuitive, 
we remark that all BTZ black hole entropy counting arguments
in the literature share this property with our proposal.

We further remark that also in the framework of the metric formulation of three-dimensional gravity
(with $AdS_3$ boundary conditions \cite{Brown:nw}), it may be difficult to account for the BTZ entropy. The connection
between the gravity formulation and our unitary treatment can be established by studying the
supergravity theory, with an ($N=2$) super-Liouville theory on the boundary
\cite{Coussaert:1995zp}\cite{Banados:1998pi}\cite{Henneaux:1999ib}, which is 
conjectured\cite{Giveon:1999zm}\cite{Giveon:1999px} / proven \cite{Hori:2001ax}\cite{Tong:2003ik}
to be dual to the $SL(2,R)/U(1)$ supercoset model\footnote{See also \cite{Pakman:2003cu}.}. 
Since the supercoset is expected to share the crucial features
of the theory we studied, it may be equally difficult to compute the (standard) black hole entropy formula
within this framework, since the effective central charge of the coset theory (or the Liouville theory)
will not be sufficiently high.

\section{Conclusions}
We conclude our long discussion with a brief summary.
In this paper, we have analyzed the Chern-Simons formulation of three-dimensional gravity.
In particular, we have mapped out the Hilbert spaces for $SL(2,R)$ Chern-Simons theory associated
with particular classical sources. We thereby clarified some of the vacuum structure of three-dimensional
gravity with a negative cosmological constant, and we made a proposal for a concrete investigation of
a quantum gravitational BPS bound.
Moreover, we have analyzed in detail the boundary conformal field theories that arise in the non-unitary
formulation, and in a minimal truncation that is unitary. We showed that the demand of a modular invariant
partition function, and unitarity leaves little room for making the BTZ black hole counting argument precise
in this context. It will be interesting to understand whether some consistency 
requirements can
be relaxed or whether different topologies can be summed over, in order 
to find an explicit model for black hole entropy counting in this simple
framework.

\section{Acknowledgements}
Thanks to all our colleagues at MIT for creating a stimulating physics environment.
 Our
research was supported by the U.S. Department of Energy under cooperative 
research agreement \# DE-FC02-94ER40818.

\appendix

\section{Conventions}
\label{conventions}
\subsection*{$SL(2,R)$}
Our conventions for $SL(2,R)$ will be (where $\sigma_i$ denote the Pauli spin matrices):
\begin{eqnarray}
T_0 &=& \frac{1}{2} (-i) \sigma_2 \nonumber \\
T_1 &=& \frac{1}{2} \sigma_3 \nonumber \\
T_2 &=& \frac{1}{2} \sigma_1 \nonumber \\
{[}T_a, T_b {]} &=& {f_{ab}}^{c} T_c
\nonumber \\
Tr(T_a T_b) &=& \frac{1}{2} \eta_{ab} \nonumber \\
\eta_{ab} &=& (-1,+1,+1) =\frac{1}{2} {f^{c}}_{ad} {f^{d}}_{bc}\nonumber \\
f_{012} &=& \epsilon_{012} =1 
\end{eqnarray}
\subsection*{Gravity}
\label{convgravity}
We define the spin connection and the
curvature two-form:
\begin{eqnarray}
\omega_{ \mu a b} &=& (e_a)^{\nu} {\nabla}_{\mu} (e_b)_{\nu} \nonumber \\
 \omega_{ab} &=& - \epsilon_{abc} \omega^c \nonumber \\
R_{ab} &=& d \omega_{ab} + \omega_{ac} {\omega^{c}}_b
\end{eqnarray}
We have the useful formula:
\begin{eqnarray}
\epsilon^{abc} R_{bc} &=& 2 d \omega^a + \epsilon^{abc} \omega_b \omega_c.
\end{eqnarray}

\section{Shifting the weight}
\label{shift}
In this appendix, we review the quantization of orbits of $SU(2)$.
We quantize a point-particle action  for a particle moving on a
quantizable orbit. In this section we closely follow \cite{Alekseev:vx},
but our analysis differs in details, and most importantly, yields a
slightly different result: the classical weight appearing in the action
is the weight of the quantum Hilbert space, shifted by half the positive
root. Technically, the difference in our approach lies in a more natural
regularization scheme. In fact, the treatment in \cite{Nielsen:1987sa} is 
precise on this point, but we believe it worthwhile to re-derive the
shift in the weight in a more fluent fashion
and in the  perhaps more familiar formalism of \cite{Alekseev:vx}.

We take over most of the conventions of \cite{Alekseev:vx}, and choose the
symplectic form on the orbit to be
\begin{eqnarray}
\Omega &=& i Tr (m \sigma^3 dg g^{-1} dg g^{-1}).
\end{eqnarray}
When we parametrize the group manifold by Euler angles as 
$g=e^{i \psi \frac{\sigma^3}{2}}e^{i \theta \frac{\sigma^2}{2}}
e^{i \phi \frac{\sigma^3}{2}} $, we can take the angles to be in
the range $\phi \in {[} 0, 2 \pi {[}$, $\theta \in {[} 0,  \pi {]}$
and  $\psi \in {[} -2 \pi, 2 \pi {[}$.
The symplectic form $\Omega$ can then be rewritten as:
\begin{eqnarray}
\Omega &=& -d (m \cos \theta) d \phi = m \sin \theta d \theta d \phi
= d \omega.
\end{eqnarray}
We can then use the differential form $\omega$ to define our particle
action:\footnote{The normalizations
 we use in the bulk of the paper for the particle
action then corresponds to identifying $\lambda= m \sigma^3$
and $m$ is half-integer.}
\begin{eqnarray}
S_p &=& \int m \cos \theta d \phi.
\end{eqnarray}

We concentrate on computing the trace of the operator
$O=e^{-m \cos \theta T}$ by performing the path integral with periodic
boundary conditions, and integrating over the boundary conditions,
while adding
the exponent of the operator $O$ to the action \cite{Alekseev:vx} to
obtain:
\begin{eqnarray}
S_{p+O} &=& \int_0^T (m \dot{\phi}-m) \cos \theta dt.
\end{eqnarray} 
We then perform the path integral over the variables
$\eta (t) = \cos \theta \in {[}-1,1{]}$ and 
$\phi (t) \in {]} - \infty ,+ \infty {[}$ with the
appropriate measure. The only subtle point is that the 
variable $\phi$ is in fact periodic, such that we have periodic boundary
conditions that include an integer winding $w \in Z$: 
\begin{eqnarray}
\phi(0) &=& \phi_i \nonumber \\
\phi(T) &=& \phi_i + 2 \pi w \nonumber.
\end{eqnarray}
Our path integral thus should incorporate a sum over winding sectors,
and we can attribute a phase 
$e^{2 \pi i w \alpha}$ to each winding 
sector (where $\alpha$ has the interpretation of a periodic $\theta$-angle
taking values in ${[} 0,1 {[}$).
Now, the only nontrivial part of the path integral is the one over
the zero-mode $\eta_0$ of $\eta$, and the sum over the winding sectors.
Following \cite{Alekseev:vx} we obtain the trace for the operator $O$:
\begin{eqnarray}
Tr (e^{-i m \cos \theta T}) 
&=& \sum_{w=-\infty}^{+\infty} \int^{1}_{-1} d \eta_0 e^{2 \pi i (m \eta_0+ \alpha)  w - i m \eta_0 T}.
\end{eqnarray} 
We use the formula:
\begin{eqnarray}
\sum_w e^{2 \pi i (m \eta_0+ \alpha)w} &=& \sum_k \delta( (m \eta_0+ \alpha)-k)
\end{eqnarray}
to perform the sum over the topological sectors. Now, we have arrived
at the essential technical point where we differ from  \cite{Alekseev:vx}.
It is clear, by using the regularization provided by the angle $\alpha$,
that we find contributions to the trace from precisely $2m$  integers $k$
(and not from $2m+1$ integers).
At this point we do need to distinguish between integer and half-integer
values for $k$, and we find that the trace of the operator is given
for integer $m$ by:
\begin{eqnarray}
Tr (e^{-i m \cos \theta T}) & = &
e^{i \alpha T-\frac{i}{2}T} 
\frac{\sin(m T)}{ \sin \frac{1}{2} T}
\end{eqnarray}
and for half-integer $m$ by:
\begin{eqnarray}
Tr (e^{-i m \cos \theta T}) & = &
e^{i \alpha T} 
\frac{\sin(m T)}{ \sin \frac{1}{2} T}
\end{eqnarray}
The most important and simple fact we take away from our analysis is that
for a given $m$ we thus obtain a representation space of
dimension $2m$ after quantization. In other words, the relation between the
spin $j$ of the representation and the weight of the orbit $m$ is given by
$2m=2j+1$, or $m= j +\frac{1}{2}$. Thus, the weight $m$ associated to the orbit
is the weight $j$ associated to the quantum Hilbert space shifted by half the
sum of the positive roots 
(which in our conventions for $SU(2)$ is $\frac{1}{2}$).
It should be clear at this stage, and from the analysis in \cite{Alekseev:vx}
that we can obtain generic matrix elements of any operator
by specifying particular initial and final conditions and performing
the path integral.

\end{document}